\documentclass[12pt]{article}
\usepackage{amsfonts,euscript}
\begin{document}

\title{A new type of supersymmetric twistors and higher spin chiral multiplets}
\author {A.A.~Zheltukhin${}^{a,b}$\\
{\normalsize ${}^a$ Kharkov Institute of Physics and Technology, 61108 Kharkov, Ukraine}\\
{\normalsize ${}^{b}$ Institute of Theoretical Physics, University of Stockholm}\\
{\normalsize  SE-10691, AlbaNova, Stockholm, Sweden}}                                            
\date{}

\maketitle

\begin{abstract}
A new type of supersymmetric twistors is proposed and they are called $\theta$-twistors versus the supertwistors. The $\theta$-twistor is a triple of spinors including the spinor superspace coordinate $\theta$ instead of the Grassmannian  scalar in the supertwistor triple. The superspace of the $\theta$-twistors is closed under the superconformal group transformations except the (super)conformal boosts. Using the $\theta$-twistors in physics preserves the auxiliary field $F$ in the chiral $(0,\frac{1}{2})$ supermultiplet  contrarily to the supertwistor description. Moreover, it yields an infinite chain of higher spin chiral supermultiplets $(\frac{1}{2},1),\, (1, \frac{3}{2}),\,(\frac{3}{2},2),...,(S, S+\frac{1}{2})$ generalizing the scalar massless supermultiplet.

\end{abstract}

\section{Introduction}
The twistor approach \cite{PR} and its supersymmetric generalization \cite{Fbr}, 
\cite{Witt1} were recently effectively applied for the calculation of multigluon amplitudes in Yang-Mills theory \cite{Witt2}, \cite{Nai}, \cite{RSV}, \cite{GKh}. An efficiency of the twistor methods in supersymmetric fields theories and superstrings appeals to further  studying  the (super)twistors and their  role in the description of superspace structures and  superfields \cite{ADHM}, \cite{CFGT}, \cite{Shir}, \cite{KS}, \cite{BC}, \cite{Ber}, \cite{Sieg}, \cite{WB}.

 Here we discuss the supertwistor origin and derive them starting from the known Cartan differential form \cite{V} in the chiral superspace. Then the  supertwistor \cite{Fbr} arises  as a triple including two commuting spinors and the  anticommuting scalar $\eta=\nu^\alpha\theta_\alpha$, 
where  $\nu^\alpha$ is commuting  Penrose spinor. Thus, the spinor superspace  coordinate $\theta$ 
  is  presented in supertwistor by its  projection breaking a  democracy among the supertwistor components. It leads to the elimination of auxiliary $F$-field from the chiral supermultiplet resulting in closure of supersymmetry transformations on the mass-shell of the spinor field \cite{Fbr}. 
Then  the problem appears how to preserve auxiliary fields during the  Penrose twistor  supersymmetrization.  This problem is studied here on the  example of  $D=4\, N=1$ supersymmetry assuming the supersymmetrization process to preserve all $\theta$ components in a twistor superspace.
To this end we introduce a new supersymmetric triple named the  $\theta$-twistor and constituted from $three$ spinors which form a nonlinear representation of the supersymmetry. 
We reveal that both the  $\theta$-twistors and supertwistors appear as the general solutions of two different supersymmetric and Lorentz covariant constraints, admissible in the chiral  superspace extended by the Penrose spinors $\nu$ and $\bar\nu$,  which generalize the standard chirality constraint. 
The symmetry properties of the $\theta$-twistor superspace are investigated and its closure under the superconformal group transformations except the (super)conformal boosts is established. 
Thereupon the Penrose contour integration is used to get the superfields independent of $\nu^\alpha$ and ${\bar\nu}^{\dot\alpha}$. These superfields describe the scalar 
supermultiplet $(0,\frac{1}{2})$ including  the  desired  $F$-field and yield an infinite chain of  massless higher spin chiral supermultiplets $(\frac{1}{2},1),\, (1, \frac{3}{2}),\,(\frac{3}{2},2),\, ......., (S, S+\frac{1}{2})$ generalizing the scalar supermultiplet. The superfields 
realize the $R$-symmetry transformations  and  may be used  for  the construction of renormalizible  Lagrangians. The studied  $D=4, \, N=1$ example is strightforward 
generalized to  the case of $SU(N)$ internal symmetry and higher dimensions $D=2,3,4(mod8)$ by analogy with the supertwistors  \cite{Fbr}, \cite{Witt1}, \cite{BZ}. 

\section{Supersymmetry }
 
In agreement with \cite{UZnc} we accept  $D=4\, N=1$ supersymmetry transformations in the  form 
\begin{equation}\label{1/1}
\begin{array}{c}
\delta\theta_\alpha=\varepsilon_\alpha,\quad 
\delta x_{\alpha\dot\alpha}=
2i(\varepsilon_{\alpha}\bar\theta_{\dot\alpha}-
\theta_{\alpha}\bar\varepsilon_{\dot\alpha}).
\end{array}
\end{equation}
The supercharges $Q_\alpha$ and $\bar Q_{\dot\alpha}$ induced  by (\ref{1/1}) are 
presented  by the differential operators 
\begin{equation}\label{2/1}
\begin{array}{c}
Q^{\alpha}=\frac{\partial}{\partial\theta_\alpha}+2i\bar\theta_{\dot\alpha}
\partial^{\dot\alpha\alpha}, \quad
{\bar Q}^{\dot\alpha}\equiv -(Q^{\alpha})^{*}=\frac{\partial}{\partial\bar\theta_{\dot\alpha}} + 2i\theta_{\alpha}\partial^{\dot\alpha\alpha},\quad
[ Q^{\alpha}, \bar Q^{\dot\alpha} ]_{+}= 4i\partial^{\dot\alpha\alpha}, 
\end{array}
\end{equation}
where $\partial^{\dot\alpha\alpha} \equiv 
\frac{\partial}{\partial x_{\alpha\dot\alpha}}$  and the correspondent supersymmetric odd derivatives $D,\bar D$ are 
\begin{equation}\label{3/1}
\begin{array}{c}
D^{\alpha}=\frac{\partial}{\partial\theta_\alpha}-2i\bar\theta_{\dot\alpha}
\partial^{\dot\alpha\alpha}, \quad
{\bar D}^{\dot\alpha}\equiv -(D^{\alpha})^{*}=\frac{\partial}{\partial\bar\theta_{\dot\alpha}}- 2i\theta_{\alpha}\partial^{\dot\alpha\alpha},
 \quad  
[ D^{\alpha},\bar D^{\dot\beta}]=-4i\partial^{\dot\beta\alpha},
\\[0.2cm]
[ Q^{\alpha}, D^{\beta}]_{+}= [ Q^{\alpha}, \bar D^{\dot\beta}]_{+}=
{[ \bar Q}^{\dot\alpha}, D^{\beta}]_{+}={ [\bar Q}^{\dot\alpha},\bar D^{\dot\beta}]_{+}=0.
\end{array}
\end{equation}
The supersymmetric even Cartan  differential one-form invariant under  (\ref{1/1}) is  
\begin{equation}\label{4/1}
\omega_{\alpha\dot\alpha}=dx_{\alpha\dot\alpha}-
2i(\theta_{\alpha}d\bar\theta_{\dot\alpha}-
d\theta_{\alpha}\bar\theta_{\dot\alpha}).
\end{equation}
A  chiral superfield $F$ is   defined  by the chirality  constraint ${\bar D}^{\dot\alpha}F=0$
which introduces  the complex chiral coordinates $y_{\alpha\dot\alpha}$
\begin{equation}\label{6/1}
y_{\alpha\dot\alpha}=x_{\alpha\dot\alpha}-2i\theta_{\alpha}\bar\theta_{\dot\alpha}
\end{equation}
satisfying the conditions 
\begin{equation}\label{7/1}
{\bar D}^{\dot\beta}y_{\alpha\dot\alpha}=0, \quad
 D^{\beta}y_{\alpha\dot\alpha}=-4i\delta^{\beta}_{\alpha}\bar\theta_{\dot\alpha}.
\end{equation}
In the comlexified superspace $(y_{\alpha\dot\alpha},\theta_{\alpha}, \bar\theta_{\dot\alpha})$  the supersymmetric derivatives and supercharges are presented in the known form \cite{WB}
\begin{equation}\label{8/1}
\begin{array}{c}
D^{\alpha}=\frac{\partial}{\partial\theta_\alpha}-
4i\bar\theta_{\dot\alpha}\partial^{\dot\alpha\alpha}, 
\quad
{\bar D}^{\dot\alpha}=\frac{\partial}{\partial\bar\theta_{\dot\alpha}},
\\[0.2cm]
Q^{\alpha}=\frac{\partial}{\partial\theta_\alpha},
\quad
{\bar Q}^{\dot\alpha}=\frac{\partial}{\partial\bar\theta_{\dot\alpha}} + 4i\theta_{\alpha}\partial^{\dot\alpha\alpha},
\end{array}
\end{equation} 
where $\partial^{\dot\alpha\alpha}=\frac{\partial}{\partial y_{\alpha\dot\alpha}}$. It yields a complex  superfield $\Phi(y_{\alpha\dot\alpha},\theta_{\alpha})$ as the general solution of the chirality constraint.
The  chiral superspace $(y_{\alpha\dot\alpha},\theta_{\alpha})$ is closed under the transformations (\ref{1/1})
\begin{equation}\label{10/1}
\delta y_{\alpha\dot\alpha}=
-4i\theta_{\alpha}\bar\varepsilon_{\dot\alpha},
\quad 
\delta\theta_\alpha=\varepsilon_\alpha
\end{equation}
which preserve the invariant chiral Cartan one-form  (\ref{4/1}) 
\begin{equation}\label{11/1}
\omega_{\alpha\dot\alpha}=dy_{\alpha\dot\alpha}+
4id\theta_{\alpha}\bar\theta_{\dot\alpha}.
\end{equation}
 The form (\ref{11/1}) is a suitable mathematical object to reveal the supertwistors introduced in \cite{Fbr}. 
 
\section{The supertwistors}

By analogy with the   Penrose twistors \cite{PR}  the  notion of the  $D=4, \, N=1$
 supertwistor \cite{Fbr} is  based on the extension of the standard  superspace  $(x_{\alpha\dot\alpha},\theta_{\alpha}, \bar\theta_{\dot\alpha})$ 
 by the even Weyl spinors $\nu_{\alpha}$ and   $\bar\nu_{\dot\alpha}$  which are inert under the supesymmetry transformations (\ref{1/1})
\begin{equation}\label{12/1}
\delta\nu_{\alpha}=0, \quad 
\delta\bar\nu_{\dot\alpha}=0 .
\end{equation} 
To introduce the supertwistors  it is suitable to start from  the complex superspace $(y_{\alpha\dot\alpha},\theta_{\alpha}, \bar\theta_{\dot\alpha})$ and the invariant Cartan form (\ref{11/1}) of which  contraction with  $\nu^{\alpha} $ 
\begin{equation}\label{13/1}
\bar s_{\dot\alpha}\equiv\nu^{\alpha}\omega_{\alpha\dot\alpha}=
d\bar q_{\dot\alpha}- d\nu^{\alpha}\bar y_{\alpha\dot\alpha}- 4i\bar\theta_{\dot\alpha}d\eta,
\end{equation} 
 yields the supersymmetric spinor one-form  $\bar s_{\dot\alpha}$. The even spinor $\bar q_{\dot\alpha}$  in  (\ref{13/1})  is  
\begin{equation}\label{14/1}
\bar q_{\dot\alpha}\equiv \nu^{\alpha}y_{\alpha\dot\alpha}
=\nu^{\alpha}x_{\alpha\dot\alpha}- 2i\eta\bar\theta_{\dot\alpha},
\end{equation} 
and the odd scalar variable $\eta$  together with  the c.c. coordinate $\bar y_{\alpha\dot\alpha}$ are defined by
\begin{equation}\label{15/1}
 \eta \equiv \nu^{\alpha}\theta_{\alpha},\quad 
\bar y_{\alpha\dot\alpha}\equiv (y_{\alpha\dot\alpha})^{*}=
x_{\alpha\dot\alpha}+2i\theta_{\alpha}\bar\theta_{\dot\alpha}.
\end{equation} 
These new composite  objects are  characterized by the relations 
\begin{equation}\label{16/1}
\bar s^{\dot\alpha}\bar s_{\dot\alpha} =\bar q^{\dot\alpha}\bar q_{\dot\alpha}=\eta^2=0.
\end{equation} 
The contraction of the one-form  $\bar s_{\dot\alpha}$  (\ref{13/1})  with 
$\bar\nu^{\dot\alpha}$ results in the supersymmetric scalar form
\begin{equation}\label{17/1}
s\equiv\bar\nu^{\dot\alpha}\bar s_{\dot\alpha}=
\bar\nu^{\dot\alpha}d\bar q_{\dot\alpha}-4i\bar\eta d\eta - d\nu^{\alpha}q_{\alpha},
\end{equation}
where $q_{\alpha}$ is the complex conjugate spinor  for $\bar q_{\dot\alpha}$
\begin{equation}\label{18/1}
q_{\alpha}\equiv(\bar q_{\dot\alpha})^*=\bar\nu^{\dot\alpha}\bar y_{\alpha\dot\alpha}=
\nu^{\alpha}x_{\alpha\dot\alpha}-2i\bar\eta\theta_{\alpha}.
\end{equation}
 We observe that the invariant differential  form $s$ (\ref{17/1})  may be rewritten as  
\begin{equation}\label{19/1}
s\equiv(\nu\omega\bar\nu)=-iZ_{\cal A}d\bar Z^{\cal A}\equiv s(Z,d\bar Z),
\end{equation}
where the triples $Z_{\cal A}$ and $\bar Z^{\cal A}$ are formed by the following composite coordinates
\begin{equation}\label{20/1}
Z_{\cal A}\equiv(-iq_{\alpha},\bar\nu^{\dot\alpha}, 2\bar\eta),
\quad
\bar Z^{\cal A}\equiv(\nu^{\alpha},i\bar q_{\dot\alpha}, 2\eta).
\end{equation}
 The triples $Z_{\cal A}$ and $\bar Z^{\cal A}$ coincide with the supetwistor and its c.c. proposed in
 \cite{Fbr} as a supersymmetric generalization of the Penrose twistors. The space of $Z_{\cal A}$ is a complex projective superspace because  of the  equivalency relation 
\begin{equation}\label{21/1}
Z_{\cal A}\sim cZ_{\cal A},
\end{equation}
where the constant $c$ is a non-zero complex number rescaling the original spinor $\nu^{\alpha}$.

Next we note   that  the differential form $s(Z,d\bar Z)$ (\ref{19/1}) may be presented as 
\begin{equation}\label{21/2}
\begin{array}{c}
is=Z_{\cal A}d\bar Z^{\cal A}= ZCdZ^{+}\equiv Z_{\cal A}C^{\cal A\cal B}(dZ_{\cal B})^{*},
\\[0.2cm]              
C^{\cal A\cal B}=
\left(
\begin{array}{lrc}0&\varepsilon_{\alpha\beta}&0\\
\varepsilon^{\dot\alpha\dot\beta}&0&0\\0&0&1\\
\end{array}
\right),
\end{array}
\end{equation}
where  $C^{\cal}$ is  treated  as a flat metric of the supertwistor space. 
We observe that $C^{\cal}$  may be used for the introduction  of a global linear form 
 (\ref{21/2})
\begin{equation}\label{21/3}
\begin{array}{c}
s(Z, \bar Z')=-\bar s(Z',\bar Z)\equiv -iZ_{\cal A}\bar Z'^{\cal A}=
\\[0.2cm]
= -q_{\alpha}\nu'^{\alpha}+ \bar\nu^{\dot\alpha}\bar q'_{\dot\alpha} -4i\bar\eta\eta',
\end{array}
\end{equation}
on the  supertwistor  space after  the  substitution of $\bar Z'^{\cal A}$ for $d\bar Z^{\cal A}$. 
The triple   $\bar Z'^{\cal A}$ is the supertwistor (\ref{20/1}),  but with $\nu'$
substituted for $\nu$
\begin{equation}\label{22/1}
\begin{array}{c}
\bar Z'^{\cal A}\equiv (\nu'^{\alpha},i\bar q'_{\dot\alpha}, 2\eta'),
\\[0.2cm]
\bar q'_{\dot\alpha}=\nu'^{\alpha}y_{\alpha\dot\alpha}, \quad \eta'=
\nu'^{\alpha}\theta_{\alpha}.
\end{array}
\end{equation}
It is  easy to check that the  scalar  form $s(Z, \bar Z')$ is a supersymmetric null form 
\begin{equation}\label{22/2}
s(Z, \bar Z')\equiv -iZ_{\cal A}\bar Z'^{\cal A}=0
\end{equation}
which  coincides  with the  bilinear form  previosly used in \cite{Fbr}. 
We see that the chiral superspace extension by an
even spinor results in the appearance of the bilinear null form $s(Z,\bar Z')$ (\ref{21/3}) associated with the triples $Z$ and $\bar Z'$ defining the supertwistor space. 
 The global form $s(Z,\bar Z')$ arises  from the scalar Cartan differential form $s(Z,d\bar Z)$ (\ref{19/1}) in the enlarged chiral superspace. It sharpens the role of chiral superspaces for the costruction of supersymmetric twistors.

\section{The superconformal symmetry }

The supertwistor components form a linear complex representation of the supersymmetry 
as it follows from the transformation relations (\ref{1/1})
\begin{equation}\label{23/1}
\delta\bar q_{\dot\alpha}=-4i\eta\bar\varepsilon_{\dot\alpha}, \quad
\delta\eta=\nu^{\alpha}\varepsilon_{\alpha}, \quad
\delta\nu^{\alpha}=0
\end{equation}
and their complex conjugate. 
The generators of  the supersymmetry  transformations (\ref{23/1}) are  given by the differential operators \begin{equation}\label{24/1}
\begin{array}{c}
Q^{\alpha}=4i\bar\eta\frac{\partial}{\partial q_{\alpha}}+
\nu^{\alpha}\frac{\partial}{\partial\eta}, \quad
{\bar Q}^{\dot\alpha}= -(Q^{\alpha})^{*}=4i\eta\frac{\partial}{\partial\bar q_{\dot\alpha}} + \bar\nu^{\dot\alpha}\frac{\partial}{\partial\bar\eta},
\\[0.2cm]
\{ Q^{\alpha}, \bar Q^{\dot\beta}\}= 4iP^{\dot\beta\alpha},\, \quad
 P^{\dot\beta\alpha} \equiv \nu^{\alpha}\frac{\partial}{\partial\bar q_{\dot\beta}} +
\bar\nu^{\dot\beta}\frac{\partial}{\partial q_{\alpha}}, 
\end{array}
\end{equation}
where $ P^{\dot\beta\alpha}$ is the hermitian  generator of translations.  
One can observe the  invariance of the bilinear form $s(Z, \bar Z')$  (\ref{21/3}) under the supersymmetry transformations (\ref{23/1}). Moreover, it was found in \cite{Fbr} that the superconformal symmetry $SU(2,2|1)$,  or $SU(2,2|N)$ for the case of $N$-extended  $D=4$ supersymmetry,  is the complete symmetry of $s(Z, \bar Z')$. To prove  this we firstly note the  symmetry in the positions $q$ and $\nu$, as well as $\eta$ and $\bar\eta$ in the bilinear form $s(Z, \bar Z')$. Then the  interchange of these coordinates in the generators (\ref{24/1}) transforms them  into the superconformal boost generators $S^{\alpha}, \,{\bar S}^{\dot\alpha}$ 
\begin{equation}\label{25/1}
S^{\alpha}=4i\eta\frac{\partial}{\partial\nu_{\alpha}}+
q^{\alpha}\frac{\partial}{\partial\bar\eta}, 
\quad
{\bar S}^{\dot\alpha}\equiv -(S^{\alpha})^{*}=4i\bar\eta\frac{\partial}{\partial\bar\nu_{\dot\alpha}} + {\bar q}^{\dot\alpha}\frac{\partial}{\partial\eta}.
\end{equation}
 The superconformal boosts act as ``square roots'' of the special conformal generator $K^{\dot\alpha\alpha}$
\begin{equation}\label{26/1}
\begin{array}{c}
\{ S^{\alpha}, \bar S^{\dot\beta}\}= 4iK^{\dot\beta\alpha},\, \quad
 K^{\dot\beta\alpha}\equiv q^{\alpha}\frac{\partial}{\partial\bar\nu_{\dot\beta}} +
{\bar q}^{\dot\beta}\frac{\partial}{\partial\nu_{\alpha}},\\[0.2cm]
[K^{\dot\alpha\alpha},  K^{\dot\beta\beta}]=\{S^{\alpha},  S^{\beta} \}=\{\bar S^{\dot\alpha},\bar S^{\dot\beta} \}=0.
\end{array}
\end{equation}
The   correspondent transformations of the supertwistor components are  given by the relations
\begin{equation}\label{27/1}
\begin{array}{c}
\delta\nu^{\alpha}=-4i\eta\xi^{\alpha}, \quad
\delta\eta={\bar q}^{\dot\alpha}\bar\xi_{\dot\alpha}, \quad
\delta\bar q_{\dot\alpha}=0,
\\[0.2cm]
\delta\nu^{\alpha}=-\kappa^{\dot\alpha\alpha}\bar q_{\dot\alpha},\quad
\delta\eta=0,\quad
\delta\bar q_{\dot\alpha}=0
\end{array}
\end{equation}
 supplemented by  their complex conjugate, where $\xi_{\alpha}$ and  $\kappa^{\dot\alpha\alpha}$ are 
  parameters of the superconformal and conformal  boosts. 
One  can check the invariance of (\ref{21/3}) under (\ref{27/1}). The commutator of the translations (\ref{24/1}) and the conformal boosts (\ref{26/1})
\begin{equation}\label{28/1}
\begin{array}{c}
 [ P^{\dot\beta\beta}, K^{\dot\alpha\alpha}]=\varepsilon^{\alpha\beta}
{\bar L}^{\dot\alpha\dot\beta} + \varepsilon^{\dot\alpha\dot\beta}L^{\alpha\beta},
\\[0.2cm]
{L}^{\alpha\beta}=q^{\alpha}\frac{\partial}{{\partial q}_{\beta}} +
 \nu^{\beta}\frac{\partial}{\partial\nu_{\alpha}},
\quad
{\bar L}^{\dot\alpha\dot\beta}={\bar q}^{\dot\alpha}\frac{\partial}{{\partial\bar q}_{\dot\beta}}
+ \bar\nu^{\dot\beta}\frac{\partial}{\partial\bar\nu_{\dot\alpha}}
\end{array}
\end{equation}
 closes  by the generators ${L}^{\alpha\beta} $ and  ${\bar L}^{\dot\alpha\dot\beta}$ satisfying by the commutation relations 
\begin{equation}\label{29/1}
\begin{array}{c}
[L^{\alpha\beta},L^{\gamma\delta}]= 
\varepsilon^{\alpha\delta}L^{\beta\gamma}- \varepsilon^{\beta\gamma}L^{\alpha\delta} ,
\\[0.2cm]
[L^{\alpha\beta}, Q^{\gamma}]=\varepsilon^{\gamma\alpha} Q^{\beta}, \quad
[{\bar L}^{\dot\alpha\dot\beta}, Q^{\gamma}]=0,
\\[0.2cm]
[{\bar L}^{\dot\alpha\dot\beta},\bar Q^{\dot\gamma}]=\varepsilon^{\dot\gamma\dot\alpha}
\bar Q^{\dot\beta},\quad
[L^{\alpha\beta},\bar Q^{\dot\gamma}]=0
\end{array}
\end{equation}
and identified with the Lorentz rotations, dilatations and phase transformations presented by  the generators $J^{\alpha\beta},\,{\bar J}^{\dot\alpha\dot\beta}$ and  $D,\,\bar D$ 
\begin{equation}\label{30/1}
\begin{array}{c}
L^{\alpha\beta}\equiv J^{\alpha\beta}+ \frac{1}{2}\varepsilon^{\alpha\beta}D,
\quad
{\bar L}^{\dot\alpha\dot\beta}\equiv {\bar J}^{\dot\alpha\dot\beta}+ \frac{1}{2}
\varepsilon^{\dot\alpha\dot\beta}{\bar D},
\\[0.2cm]
 J^{\alpha\beta}\equiv \frac{1}{2}(L^{\alpha\beta} + L^{\beta\alpha}), 
\quad
D\equiv q_{\alpha}\frac{\partial}{{\partial q}_{\alpha}} -\nu_{\alpha}\frac{\partial}{\partial\nu_{\alpha}},
\\[0.2cm]
{\bar L}^{\dot\alpha\dot\beta}\equiv \frac{1}{2}({\bar L}^{\dot\alpha\dot\beta} + {\bar L}^{\dot\beta\dot\alpha}),\quad
{\bar D}\equiv {\bar q}_{\dot\alpha}\frac{\partial}{\partial{\bar q}_{\dot\alpha}} -\bar\nu_{\dot\alpha}\frac{\partial}{\partial\bar\nu_{\dot\alpha}}\quad
\end{array}
\end{equation}
which are the symmetries of the  bilinear form too. 
 The dilatation $D_R$ and  the phase rotation  $D_I$ generators are presented by the hermitian combinations of $D$ and $\bar D$
\begin{equation}\label{30/2}
D_{R}= (D+\bar D), \quad D_{I}=i(D-\bar D)
\end{equation}
and produce correspondent  finite transformations of the supertwistor components
\begin{equation}\label{31/1}
\begin{array}{c}
q'_{\beta}=e^{\varphi}q_{\beta}, 
\quad
{\bar q}'_{\dot\beta}=e^{\varphi*}{\bar q}_{\dot\beta},
 \quad
\nu'_{\beta}=e^{-\varphi}\nu_{\beta} ,
\quad 
\bar\nu'_{\dot\beta}=e^{-\varphi*}\bar\nu_{\dot\beta},
\end{array}
\end{equation}
where $\varphi=\varphi_{R}+i\varphi_{I}$ is  a complex parameter. Eqs. (\ref{31/1}) results in the relation 
\begin{equation}\label{31/2}
\begin{array}{c}
Z'_{\cal A}= e^{i\varphi_{I}}(-iq_{\alpha}e^{\varphi_{R}} \,,\,
\bar\nu^{\dot\alpha}e^{-\varphi_{R}}\,,\, 2\bar\eta e^{-i\varphi_{I}}), \quad
\bar Z'^{\cal A}= 
e^{-i\varphi_{I}}(\nu_{\alpha}e^{-\varphi_{R}}\, ,\,i{\bar q}_{\dot\alpha}e^{\varphi_{R}}\,,\, 2\eta e^{i\varphi_{I}} ).
\end{array}
\end{equation}
proving the  invariance of the bilinear form (\ref{21/3}). The phase symmetry turns out to be trivial, because of the projectivity  of the supertwistor  space. To prove this observation we note that 
the anticommutator of the supercharge with the  superconformal boost
\begin{equation}\label{32/1}
\begin{array}{c}
\{ Q^{\alpha}, {\bar S}^{\dot\beta}\}=0,
\\[0.2cm]
\{ Q^{\alpha}, S^{\beta} \}=4iL^{\beta\alpha} +4\varepsilon^{\alpha\beta}U_{5}
\end{array}
\end{equation}
yields one additional symmetry of the bilinear form described by the  hermitian generator $U_{5}$ \begin{equation}\label{32/2}
U_{5}=i(\eta\frac{\partial}{\partial\eta} - \bar\eta\frac{\partial}{\partial\bar\eta})
\end{equation}
which produces the phase rotations of the odd variables $\eta$ and  $\bar\eta$  
\begin{equation}\label{33/1}
\begin{array}{c}
\eta'=e^{i\lambda}\eta, \quad
\bar\eta'=e^{-i\lambda}\bar\eta_.
\end{array}
\end{equation}
The generator $U_{5}$ has non-zero commutation relations only with $Q^{\beta}, S^{\beta}$ and their c.c.
\begin{equation}\label{33/2}
\begin{array}{c}
[U_{5}, Q^{\beta}]=-iQ^{\beta},  \quad [U_{5}, {\bar Q}^{\dot\beta}]=i{\bar Q}^{\dot\beta}, 
\\[0.2cm]
[U_{5}, S^{\beta}]=iS^{\beta},  \quad [U_{5}, {\bar S}^{\dot\beta}]=-i{\bar S}^{\dot\beta}
\end{array}
\end{equation}
and its appearance  does not break the closure of superalgebra preserving the twistor bilinear form. 
Using the  $U_{5}$  symmetry  may compensate the phase factor in front of $\eta $ and  $\bar\eta$ components of $Z'^{\cal A}$  and  $\bar Z'^{\cal A}$ in (\ref{31/2}) resulting in 
\begin{equation}\label{33/3}
Z'_{\cal A}=e^{i\varphi_{I}}
Z_{\cal A}, \quad 
{\bar Z}'^{\cal A}=e^{-i\varphi_{I}}
{\bar Z}^{\cal A},
\end{equation}
where  the dilatation parameter $\varphi_{R}=0$. 
This shows that the effect of the  phase transformation associated with the generator $D_{I}$ is equvalent to the multiplication of the supertwistor $Z$ and $\bar Z$ by a  phase factor.  
But, the $ Z$-space is a projective space and it means that the phase symmetry  (\ref{31/1}) is  excluded  from the  list of the  bilinear form symmetries. 

So, we get  15 real symmetries associated with the bosonic sector of  the  supertwistor space:\, $ 4(P) + 4(K) +6(J, \bar J) + 1 (D_{R})=15$. 
Next we observe that the pure  phase part of  (\ref{31/1}) can be treated as the axial or $\gamma_{5}$ rotation of  $\theta_{\alpha}$  and  ${\bar\theta}_{\alpha}$ 
\begin{equation}\label{34/1}
\theta'_{\beta}=e^{ia_{5}}\theta_{\beta}, \quad 
\bar\theta'_{\dot\beta}=e^{-ia_{5}}\bar\theta_{\dot\beta}. 
\end{equation}
 In the addition to the above mentioned 15 bosonic symmetries there are 8 real symmetries associated with the fermionic sector of the supertwistor space:\,  $  4(Q, \bar Q) + 4(S, \bar S)=8$, that 
 complete the list of symmetries of supeconformal group $SU(2,2|1)$. 

Let us remark that the action of the superconformal  and  conformal  symmetries on the coordinates of the chiral superspace ($y_{\beta\dot\beta}, \theta_{\beta}$) is realized by the  nonlinear transformations  
\begin{equation}\label{36/1}
\begin{array}{c}
\delta y_{\alpha\dot\alpha}\equiv(\xi_{\alpha}S^{\alpha}+ \bar\xi_{\dot\alpha}{\bar S}^{\dot\alpha})y_{\alpha\dot\alpha}=4i\theta_{\alpha}(\xi^{\beta}y_{\beta\dot\alpha}),\quad 
\delta\theta_{\alpha}=y_{\alpha\dot\beta}\bar\xi^{\dot\beta} -4i\theta_{\alpha}(\xi^{\beta}\theta_{\beta}),
\\[0.2cm]
\delta y_{\alpha\dot\alpha}\equiv(\kappa_{\beta\dot\beta}K^{\dot\beta\beta}) y_{\alpha\dot\alpha}= -(y\kappa y)_{\alpha\dot\alpha},  
\quad 
\delta\theta_{\alpha}=-(y\kappa\theta)_{\alpha},
\end{array}
\end{equation}
where 
$(y\kappa y)_{\alpha\dot\alpha}=y_{\alpha\dot\beta}\kappa^{\dot\beta\beta}y_{\beta\dot\alpha}$ and   
$(y\kappa y)_{\alpha\dot\alpha}=-y_{\alpha\dot\beta}\kappa^{\dot\beta\beta}\theta_{\beta}$,  and  $ S , \bar S,  K $ are the (super)conformal  generators. 
So, we conclude that the extension of the chiral superspace by  $\nu$ and $\bar\nu$ transforms   the nonlinear action  of the supeconformal group to the linear action  realized by the  composite coordinates in the enlarged chiral superspace belonging to the  supertwistor componets. 

In the next section we present another  composite triple in the enlarged chiral superspace.

\section{Supersymmetric $\theta$-twistors versus supertwistors}

Having a well defined base forthe supertwistor conception we are ready now to introduce 
 an alternative supersymmetric triple in the enlarged chiral superspace who we shall call $\theta$-twistor and that realizes a nonlinear representation of the supersymmetry.
To this end we observe that the spinor differential one-form  $s_{\alpha}$  which is complex conjugate of ${\bar s}_{\dot\alpha}$ (\ref{13/1})
\begin{equation}\label{37/1}
s_{\alpha}= ({\bar s}_{\dot\alpha})^* =
 \omega_{\alpha\dot\alpha}\bar\nu^{\dot\alpha}=
d q_{\alpha}- y_{\alpha\dot\alpha}d\bar\nu^{\dot\alpha}- 4i\theta_{\alpha}d\bar\eta,
\end{equation}
may be  presented in the equivalent form
\begin{equation}\label{38/1}
s_{\alpha}=dl_{\alpha}-  y_{\alpha\dot\alpha}d\bar\nu^{\dot\alpha}+ 
4id\theta_{\alpha}\bar\eta,
\end{equation} 
where the  composite even spinor $l_{\alpha}$   is defined  by the relation
\begin{equation}\label{39/1}
l_{\alpha}\equiv y_{\alpha\dot\alpha}\bar\nu^{\dot\alpha}
=x_{\alpha\dot\alpha}\bar\nu^{\dot\alpha}- 2i\theta_{\alpha}\bar\eta .
\end{equation} 
 The spinor $l_{\alpha}$ is connected with the spinor supertwistor component $q_{\alpha}$ (\ref{18/1})
by the  shift 
\begin{equation}\label{40/1}
l_{\alpha}=q_{\alpha}- 4i\theta_{\alpha}\bar\eta.
\end{equation}
The shift provides new transformation properties of $l_{\alpha}$ under the supersymmetry 
\begin{equation}\label{41/1}
{\delta l}_{\alpha}=-4i\theta_{\alpha}(\bar\nu^{\dot\beta}\bar\varepsilon_{\dot\beta}), \quad
\delta\theta_{\alpha}=\varepsilon_{\alpha}, \quad
\delta\bar\nu_{\dot\alpha}=0
\end{equation}
supplemented  by their c.c. relations. In contrast to the linear trasformations (\ref{23/1}), realized by the supertwistor componets,  the new triple $\Xi_{\cal A}$ and its c.c.  $\bar\Xi^{\cal A}$ 
\begin{equation}\label{42/1}
\Xi_{\cal A}\equiv(-il_{\alpha},\bar\nu^{\dot\alpha},\theta^{\alpha}),
\quad
\bar\Xi^{\cal A}\equiv(\nu^{\alpha},i{\bar l}_{\dot\alpha},\bar\theta^{\dot\alpha})
\end{equation} 
yield a  nonlinear representations of the supersymmetry.
 One  can see that the triple $\Xi_{\cal A}$ is obtained from the supertwistor $Z_{\cal A}$ by the substitution $q_{\alpha}\rightarrow l_{\alpha} $ and $2\bar\eta\rightarrow\theta^{\alpha}$. Then the substitution of $l_{\alpha}$ for $q_{\alpha}$ in the linear form (\ref{21/3}) with using Eq. (\ref{40/1}) results in the relation 
\begin{equation}\label{43/1}
s(Z,\bar Z)\equiv
-iZ_{\cal A}\bar Z'^{\cal A}=-i\Xi_{\cal A}\circ\bar\Xi'^{\cal A}\equiv{\tilde s}(\Xi,\bar\Xi'),
\end{equation}
 where the nonlinear form ${\tilde s}(\Xi,\bar\Xi')$ in the $\Xi$-triple superspace (\ref{42/1}) is defined as 
\begin{equation}\label{44/1}
\tilde s
(\Xi,\bar\Xi')=
-i\Xi_{\cal A}\circ\bar\Xi'^{\cal A}
\equiv  -l_{\alpha}\nu'^{\alpha}+ 
\bar\nu^{\dot\alpha}\bar l'_{\dot\alpha} - 4i\eta'\bar\eta.
\end{equation} 
The  relation (\ref{44/1}) differs from (\ref{21/3}) by the changes of the sign in front of the term $\eta'\bar\eta$ and the substitution of $l$ for  $q $. The form  (\ref{44/1}) becomes a nonlinear in the $\Xi$-triple superspace (\ref{42/1})
 \begin{equation}\label{44/2}
\tilde s
(\Xi,\bar\Xi')=
-i\Xi_{\cal A}\circ
\bar\Xi'^{\cal A}
\equiv  -l_{\alpha}\nu'^{\alpha}+ 
\bar\nu^{\dot\alpha}\bar l'_{\dot\alpha} -i
g_{\alpha\dot\alpha}\theta^{\alpha}\bar\theta^{\dot\alpha}=0,\quad 
g_{\alpha\dot\alpha}\equiv 
4\nu'_{\alpha}\bar\nu_{\dot\alpha},
\end{equation}
where we observe  the  ``metric'' factor  $g_{\alpha\dot\alpha}$ appearance  in front of the odd component contribution. 
By analogy with (\ref{21/2}) the nonlinear form  $\Xi_{\cal A}\circ\bar\Xi'^{\cal A}$ may be rewritten as
\begin{equation}\label{45/1}
\begin{array}{c}
\Xi_{\cal A}\circ\bar\Xi'^{\cal A}= \Xi G \Xi'^{+}\equiv \Xi_{\cal A}G^{\cal A\cal B}(\Xi'_{\cal B})^{*},\\[0.2cm]
G^{\cal A\cal B}=
\left(
\begin{array}{lrc}0&\varepsilon_{\alpha\beta}&0\\
\varepsilon^{\dot\alpha\dot\beta}&0&0\\0&0&4\nu'_{\alpha}\bar\nu_{\dot\beta}\\
\end{array}
\right),
\end{array}
\end{equation}
where $G^{\cal}$ plays  the role of a non-flat  ``metric''  in the $\Xi$-triple space. 
Trying to consider (\ref{45/1}) as a natuaral nonlinear form in the  $\Xi$-superspace one could resume  that the transition to $\Xi$-triples might be interpreted as a curving the supertwistor space described  by $g_{\alpha\dot\alpha}$. This effective interpretation is partially supported by the  relation  between the supersymmetric one-form (\ref{19/1}) and the nonlinear  differential form $\Xi G d\Xi^{+}$
\begin{equation}\label{46/1}
\Xi G d\Xi^{+}=ZC dZ^{+}  -  4(d\nu^{\alpha}\theta_{\alpha})\bar\eta
\end{equation}
which  may be rewritten  in an equivalent form as
\begin{equation}\label{47/1}
\begin{array}{c}
ZC dZ^{+}= \Xi (G d +A)\Xi^{+}, \quad
A= \left(
\begin{array}{lrc}0&0&0\\
0&0&0\\0&0&4d\nu_{\alpha}\bar\nu_{\dot\beta}\\
\end{array}
\right).
\end{array}
\end{equation}
This representation   introduces the  ``connection''   $A$ needed for the construction of the covariant differential in the ``curved''   $\Xi$-superspace.
 Another remark is that the odd components  of the  $\theta$-twistors (\ref{42/1}) break the projective character of its  bosonic  components. In this connection 
it is important  to know  whether the superconformal symmetry survives the transition to the $\theta$-twistors from the supertwistors.  We shall consider this problem below.

\section{Symmetries of the $\theta$-twistor superspace}

To study the properties  of the $\Xi$-space under the superconformal symmetry we need  to present its  generators in terms of the $\theta$-twistor  components.
The generators of the supersymmetry transformations (\ref{41/1})  just have the requred form 
\begin{equation}\label{48/1}
\begin{array}{c}
Q^{\alpha}=\frac{\partial}{\partial\theta_{\alpha}}+4i\nu^{\alpha}(\bar\theta_{\dot\beta}\frac{\partial}{\partial\bar l_{\dot\beta}}),
 \quad
 \bar Q^{\dot\alpha}\equiv -(Q^{\alpha})^*=\frac{\partial}{\partial\bar\theta_{\dot\alpha}}+ 4i\bar\nu^{\dot\alpha}(\theta_{\beta}\frac{\partial}{\partial l_{\beta}})
\end{array}
\end{equation}
with the  anticommutator of  $Q$ and $\bar Q$ closed  by the translation generator $ P^{\dot\alpha\alpha}$ 
\begin{equation}\label{49/1}
\begin{array}{c}
\{ Q^{\alpha}, \bar Q^{\dot\beta}\}= 4iP^{\dot\beta\alpha},\, \quad
 P^{\dot\beta\alpha} \equiv (\bar\nu^{\dot\beta}\frac{\partial}{\partial l_{\alpha}}+
\nu^{\alpha}\frac{\partial}{\partial\bar l_{\dot\beta}}),
\\[0.2cm]
\{ Q^{\gamma},P^{\dot\beta\alpha}\}=0, \quad \{\bar Q^{\dot\gamma},P^{\dot\beta\alpha}\}=0.    
\end{array}
\end{equation}
The generators (\ref{49/1}) preserve the  Lorentz invariant form (\ref{45/1}) and transform the 
$\Xi$-space into itself. The  Lorentz generators  $J^{\alpha\beta},\,{\bar J}^{\dot\alpha\dot\beta}$ constructed  from the $\theta$-twistor components are presented by the  symmetric combinations  of the  new differential operators  ${L}^{\alpha\beta},\, {\bar L}^{\dot\alpha\dot\beta}$ generalizing the supertwistor generators (\ref{28/1})
\begin{equation}\label{50/1}
\begin{array}{c}
{L}^{\alpha\beta}=l^{\alpha}\frac{\partial}{{\partial l}_{\beta}} +
 \nu^{\beta}\frac{\partial}{\partial\nu_{\alpha}}+ \theta^{\alpha}\frac{\partial}{\partial\theta_{\beta}},
\\[0.2cm]
{\bar L}^{\dot\alpha\dot\beta}={\bar l}^{\dot\alpha}\frac{\partial}{{\partial\bar l}_{\dot\beta}}
+ \bar\nu^{\dot\beta}\frac{\partial}{\partial\bar\nu_{\dot\alpha}}+ \bar\theta^{\dot\alpha}\frac{\partial}{\partial\bar\theta_{\dot\beta}},
\\[0.2cm]
[L^{\alpha\beta},L^{\gamma\delta}]=
\varepsilon^{\alpha\delta}L^{\beta\gamma}-\varepsilon^{\beta\gamma}L^{\alpha\delta} ,
\\[0.2cm]
[L^{\alpha\beta}, Q^{\gamma}]= \varepsilon^{\gamma\alpha} Q^{\beta},\quad 
[L^{\alpha\beta},\bar Q^{\dot\gamma}]=0,
\\[0.2cm]
[L^{\alpha\beta}, P^{\dot\gamma\gamma}]= \varepsilon^{\gamma\alpha} P^{\dot\gamma\beta},\quad 
[\bar L^{\dot\alpha\dot\beta}, P^{\dot\gamma\gamma}]=\varepsilon^{\dot\gamma\dot\alpha} P^{\dot\beta\gamma}.
\end{array}
\end{equation}
 
Another irreducible combinations  of ${L}^{\alpha\beta},\, {\bar L}^{\dot\alpha\dot\beta}$ yield the scale and the phase transformation generators  $D,\,\bar D$ that  together  with   $J^{\alpha\beta}, {\bar J}^{\dot\alpha\dot\beta}$ are given by 
\begin{equation}\label{50/2}
\begin{array}{c}
L^{\alpha\beta}\equiv J^{\alpha\beta}+ \frac{1}{2}\varepsilon^{\alpha\beta}D,
\quad
{\bar L}^{\dot\alpha\dot\beta}\equiv {\bar J}^{\dot\alpha\dot\beta}+ \frac{1}{2}\varepsilon^{\dot\alpha\dot\beta}{\bar D},
\\[0.2cm]
 J^{\alpha\beta}\equiv \frac{1}{2}(L^{\alpha\beta} + L^{\beta\alpha}), 
\quad
D\equiv l_{\alpha}\frac{\partial}{{\partial l}_{\alpha}} -\nu_{\alpha}\frac{\partial}{\partial\nu_{\alpha}}+  \theta_{\alpha}\frac{\partial}{\partial\theta_{\alpha}}  ,
\\[0.2cm]
{\bar L}^{\dot\alpha\dot\beta}\equiv \frac{1}{2}({\bar L}^{\dot\alpha\dot\beta} + {\bar L}^{\dot\beta\dot\alpha}),\quad
{\bar D}\equiv {\bar l}_{\dot\alpha}\frac{\partial}{\partial{\bar l}_{\dot\alpha}} -\bar\nu_{\dot\alpha}\frac{\partial}{\partial\bar\nu_{\dot\alpha}}+ \bar\theta_{\dot\alpha}\frac{\partial}{\partial\bar\theta_{\dot\alpha}}
\end{array}
\end{equation}
and contain only  the $\theta$-twistor components. 
It is  easy to check that $J^{\alpha\beta}, {\bar J}^{\dot\alpha\dot\beta}$ have  the standard  commutators with  the  supercharges (\ref{48/1})
\begin{equation}\label{51/1}
\begin{array}{c}
[J^{\alpha\beta}, Q^{\gamma}]=\frac{1}{2}(\varepsilon^{\gamma\alpha} Q^{\beta}+
\varepsilon^{\gamma\beta} Q^{\alpha}), \quad
[{\bar J}^{\dot\alpha\dot\beta}, Q^{\gamma}]=0,
\\[0.2cm]
[{\bar J}^{\dot\alpha\dot\beta},\bar Q^{\dot\gamma}]=\frac{1}{2}(\varepsilon^{\dot\gamma\dot\alpha}\bar Q^{\dot\beta} +\varepsilon^{\dot\gamma\dot\beta}\bar Q^{\dot\alpha}), \quad
[ J^{\alpha\beta}, \bar Q^{\dot\gamma}]=0.
\end{array}
\end{equation} 
 and with $P^{\dot\alpha\alpha}$ (\ref{48/1}) showing a realizability of the super Poincare algebra by  the $\theta$-twistors.
The dilatation $D_R$ and the  phase rotation  $D_I$ generators  are  formed by the  hermitian combinations of $D$ and $\bar D$
\begin{equation}\label{52/1}
D_{R}=(D+\bar D), \quad D_{I}= i(D-\bar D).
\end{equation}
The finite transformations of  the  $\theta$-twistor components generated by  $D$  and $\bar D$ are
\begin{equation}\label{52/2}
\begin{array}{c}
l'_{\beta}=e^{\varphi}l_{\beta}, 
\quad
{\bar l}'_{\dot\beta}=e^{\varphi*}{\bar l}_{\dot\beta},
 \quad
\nu'_{\beta}=e^{-\varphi}\nu_{\beta} ,
\quad 
\bar\nu'_{\dot\beta}=e^{-\varphi*}\bar\nu_{\dot\beta},
\\[0.2cm]
\theta'_{\beta}=e^{\varphi}\theta_{\beta}, 
\quad
\bar\theta'_{\dot\beta}=e^{\varphi*}\bar\theta_{\dot\beta}
\end{array}
\end{equation}
or equivalently 
\begin{equation}\label{52/3}
\begin{array}{c}
\Xi'_{\cal A}= e^{\varphi}(-il_{\alpha} \,,\,
\bar\nu^{\dot\alpha}e^{-2\varphi_{R}}\,,\, \theta^{\alpha}), \quad
\bar\Xi'^{\cal A}= e^{\varphi*}(\nu_{\alpha}e^{-2\varphi_{R}}\, ,\,i{\bar l}_{\dot\alpha}\,,\, \bar\theta^{\dot\alpha}).
\end{array}
\end{equation}
It yields the transformation of the form $i{\tilde s}(\Xi,\bar{\tilde\Xi})$ spanned on the $\theta$-twistors $\Xi$ and  $\bar{\tilde\Xi}$
 \begin{equation}\label{53/1}
\begin{array}{c}
\Xi'\circ\bar{\tilde\Xi'}^{\cal A}= e^{2\varphi_{R}}(-il_{\alpha} \,,\,
\bar\nu^{\dot\alpha}e^{-2\varphi_{R}}\,,\, \theta^{\alpha})
({\tilde\nu}^{\alpha}e^{-2\varphi_{R}}\, ,\,i{\bar{\tilde l}}_{\dot\alpha}\,,\, \bar{\tilde\theta}^{\dot\alpha})
\\[0.2cm]
=\Xi\circ\bar{\tilde\Xi}^{\cal A}+4i( e^{-2\varphi_{R}}-1)\tilde\eta\bar\eta
\end{array}
\end{equation}
 which shows that  the form (\ref{45/1}) is not invariant under  the dilatations 
(\ref{52/1}) generated  by $D_{R}$.
From the other hand the phase transformations generated by $D_{I}$ 
preserve  the form (\ref{45/1}),  because they yield  multiplication of $\Xi$ and $\bar{\tilde\Xi}$ by the  phase factors 
\begin{equation}\label{54/1}
\Xi'_{\cal A}=
e^{i\varphi_{I}}
\Xi_{\cal A}, \quad 
{\bar{\tilde\Xi'}}^{\cal A}=
e^{-i\varphi_{I}}
{\bar{\tilde\Xi}}^{\cal A},
\end{equation}
as it follows from (\ref{53/1}) with  $\varphi_{R}=0$.
As the $\Xi$-triple space don't form  projective superspace
 the phase transformation  (\ref{54/1}) is  not a trivial symmetry of  the $\Xi$-space.
In spite of the scale symmetry breaking for the nonlinear form (\ref{53/1}) the dilatations transform  the $\theta$-twistor space into itself. 
Moreover, the generators $D_{R}$ and $D_{I}$ yield the proper commutators with  the  supercharges  $Q^{\alpha},\bar Q^{\dot\alpha}$ (\ref{48/1}) and $P^{\dot\alpha\alpha}$ (\ref{49/1})
\begin{equation}\label{54/2}
\begin{array}{c}
[D,\, Q^{\alpha}]= -Q^{\alpha}, \quad [ D,\, \bar Q^{\dot\alpha}]=0,\quad
[D,\,P^{\dot\alpha\alpha}]=-P^{\dot\alpha\alpha},
\\[0.2cm]
[\bar D,\, \bar Q^{\dot\alpha}]= -\bar Q^{\dot\alpha},\quad[\bar D,\, Q^{\alpha}]=0,\quad
[\bar D,\,P^{\dot\alpha\alpha}]=-P^{\dot\alpha\alpha}
\end{array}
\end{equation}
proving the closure of the superalgebra realized by the $\theta$-twistors.  

Similarly to the supertwistors the  form (\ref{45/1}) is invariant under the $\gamma_5$ rotations  (\ref{34/1}) carrying  out  phase rotations of $\theta$ and  $\bar\theta$ 
\begin{equation}\label{55/1}
 U_{5}= i(\theta_{\alpha}\frac{\partial}{\partial\theta_{\alpha}}- \bar\theta_{\dot\alpha}\frac{\partial}{\partial\bar\theta_{\dot\alpha}}).
\end{equation}
The phase rotation generator  $U_{5}$ coincides with the $\theta$ dependent part of the generator 
$D_{I}$ (\ref{51/1})
 and it  has proper commutation relations with  $Q^{\beta}, \bar Q^{\dot\alpha}$ (\ref{48/1}) and ${\bar J}^{\dot\alpha\dot\beta}$  (\ref{50/2})
\begin{equation}\label{56/1}
\begin{array}{c}
[U_{5},\, Q^{\alpha}]=-iQ^{\alpha},  \quad 
[U_{5}, \,{\bar Q}^{\dot\alpha}]=i{\bar Q}^{\dot\alpha}, \quad  [U_{5},\, P^{\dot\alpha\alpha}]=0,
\\[0.2cm]
[U_{5},\ L^{\alpha\beta}]=0, \quad 
[U_{5}, \,{\bar L}^{\dot\alpha\dot\beta}]=0.
\end{array}
\end{equation}

Thus, the $\Xi$-space turns out to be  invariant under the super Poincare algebra, scaling and phase transformations (\ref{52/2}) accompanied by  the axial  rotations  (\ref{55/1}).

Next is to check whether the (super)conformal boosts are symmetry of  the $\theta$-twistor space.
 An  obstacle is the $\theta$-dependent term presence in the Lorentz and dilation generators (\ref{50/2}). The conformal boost (\ref{26/1}) of supertwistors was deduced from  $P^{\dot\beta\alpha}$ by the substitution of $q_{\alpha}$  for $\nu_{\alpha}$. Repetition of this  trick anew does not provide the proper r.h.s. of the commutator  $[P^{\dot\beta\beta}, K^{\dot\alpha\alpha}]$, because of  loss of the $\theta$-contribution  in  the Lorentz and dilatation generators. 
So, we have to seek $K^{\dot\alpha\alpha}$ in a form of general $\Xi,\,\bar\Xi$ dependent hermitian differential  operator
\begin{equation}\label{57/1}
\begin{array}{c}
 K^{\dot\alpha\alpha}=( \,
 F^{\dot\alpha\alpha}_{\gamma}\frac{\partial}{\partial\nu_{\gamma}}+
{\bar F}^{\dot\alpha\alpha}_{\dot\gamma}\frac{\partial}{\partial\bar\nu_{\dot\gamma}}\, )+
( \, G^{\dot\alpha\alpha}_{\gamma}\frac{\partial}{\partial l_{\gamma}}
+{\bar G}^{\dot\alpha\alpha}_{\dot\gamma}\frac{\partial}{\partial\bar l_{\dot\gamma}}\, )
+  ( \,\Psi^{\dot\alpha\alpha}_{\gamma}\frac{\partial}{\partial\theta_{\gamma}}+
{\bar\Psi}^{\dot\alpha\alpha}_{\dot\gamma}\frac{\partial}{\partial\bar\theta_{\dot\gamma}}\, ).
\end{array}
\end{equation}
The unknown complex  functions  
$ F= F(\Xi, \bar\Xi)$, $ G= G(\Xi, \bar\Xi)$,  $\Psi= \Psi(\Xi, \bar\Xi)$  and  their c.c. $\bar F, \bar G, \bar\Psi $ in (\ref{57/1}) have to be fixed by the superconformal algebra with  the 
$[K,\ P]$ commutator fixed  by 
\begin{equation}\label{58/1}
[ P^{\dot\beta\beta}, K^{\dot\alpha\alpha}]=\varepsilon^{\alpha\beta}
{\bar L}^{\dot\alpha\dot\beta} + \varepsilon^{\dot\alpha\dot\beta}L^{\alpha\beta}.
\end{equation}
 The substitution of (\ref{57/1}) into (\ref{58/1}) gives the contribution of the terms linear in  $\frac{\partial}{\partial\theta_{\gamma}}$  and $\frac{\partial}{\partial\bar\theta_{\dot\gamma}}$ 
\begin{equation}\label{59/1}
[ P^{\dot\beta\beta},  \,\Psi^{\dot\alpha\alpha}_{\gamma}\frac{\partial}{\partial\theta_{\gamma}}+  \,
{\bar\Psi}^{\dot\alpha\alpha}_{\dot\gamma}\frac{\partial}{\partial\bar\theta_{\dot\gamma}}]= 
[ P^{\dot\beta\beta},  \, \Psi^{\dot\alpha\alpha}_{\gamma}]\frac{\partial}{\partial\theta_{\gamma}} +  \,
[ P^{\dot\beta\beta},  \, {\bar\Psi}^{\dot\alpha\alpha}_{\dot\gamma}]\frac{\partial}{\partial\bar\theta_{\dot\gamma}} 
\end{equation}  
 and next using $ L^{\alpha\beta},\, {\bar L}^{\dot\alpha\dot\beta}$ (\ref{50/1}) and  $P^{\dot\beta\beta}$  (\ref{49/1}) transforms Eq. (\ref{58/1}) to the equation
\begin{equation}\label{60/1}
( \,\nu^{\beta}\frac{\partial}{\partial\bar l_{\dot\beta}} + 
\bar\nu^{\dot\beta}\frac{\partial}{\partial l_{\beta}} \,)
\Psi^{\dot\alpha\alpha}_{\gamma}\frac{\partial}{\partial\theta_{\gamma}} 
+
( \,\nu^{\beta}\frac{\partial}{\partial\bar l_{\dot\beta}}+ \bar\nu^{\dot\beta}\frac{\partial}{\partial l_{\beta}\,}){\bar\Psi}^{\dot\alpha\alpha}_{\dot\gamma}\frac{\partial}{\partial\bar\theta_{\dot\gamma}}= 
\varepsilon^{\dot\alpha\dot\beta}
\theta^{\alpha}\frac{\partial}{\partial\theta_{\beta}} + 
\varepsilon^{\alpha\beta}\bar\theta^{\dot\alpha}\frac{\partial}{\partial\bar\theta_{\dot\beta}}\,\,.
\end{equation}
  It is easy to see the inconsistency of Eq. (\ref{60/1}) due to the dependence  of its  l.h.s. on the even spinors $\nu_{\beta}, \,\bar\nu_{\dot\beta}$ contrarily to its r.h.s..  This $\nu$-dependence  can't be removed because of the identities $\nu_{\beta}\nu^{\beta}= \bar\nu_{\dot\beta}\bar\nu^{\dot\beta}=0$ and impossibility to construct antisymmetric tensors in the l.h.s. of (\ref{60/1}) which might absorb $\nu,\bar\nu$. So, we run into the problem of the conformal boost generator construction by the $\theta$-twistors resulting in  the superconformal symmetry breakdown up to its subgroup including the super Poincare, scaling and axial symmetries. 

\section{Chiral supermultiplets of higher spin fields}

As it was above discussed the supertwistor appears as a natural object in the chiral coordinate superspace $ (y_{\alpha\dot\alpha},\,\theta_{\alpha},\, \bar\theta_{\dot\alpha})$ extended by the even Majorana spinor  $(\nu_{\alpha},\, \bar\nu^{\dot\alpha})$.

  But, one can arrive to the supertwistors starting from the supersymmetric constraints 
\begin{equation}\label{61/1}
\begin{array}{c}
{\bar D}^{\dot\alpha}F(x,\theta,\bar\theta)=0 \longrightarrow \, 
F=F(y,\theta),
\\[0.2cm]
\nu_{\alpha}D^{\alpha}F(y,\theta,\nu)=0  \longrightarrow \, F=F(\bar Z^{\cal A}),
\end{array}
\end{equation}
 whose  general solution is a chiral superfield $F(\bar Z^{\cal A})$ depending  just on the supertwistor  $\bar Z^{\cal A}$. It is a cosequence of  the observation that 
 $\bar Z^{\cal A}\equiv(\nu^{\alpha},i\bar q_{\dot\alpha}, 2\eta)$ is the general solution of the second constraint in Eqs. (\ref{61/1})
\begin{equation}\label{62/1}
\begin{array}{c}
\nu_{\alpha}D^{\alpha}\bar Z^{\cal A}=0
\end{array}
\end{equation}
as it follows from Eqs. (\ref{7/1}) and (\ref{14/1}) and the relations  
\begin{equation}\label{63/1}
{\bar D}^{\dot\alpha}\bar q_{\dot\beta}={\bar D}^{\dot\alpha}\eta=0, \quad  
D^{\alpha}\bar q_{\dot\beta}=-4i\nu^{\alpha}\bar\theta_{\dot\beta},\quad D^{\alpha}\eta=\nu^{\alpha}, 
\quad  
\frac{\partial}{\partial y_{\alpha\dot\alpha}}\bar q_{\dot\beta}= \delta^{\dot\alpha}_{\dot\beta}
\nu_{\alpha}.
\end{equation}

But, if we choose other supersymmetric constraints including $\bar\nu$ instead of $\nu$ as in (\ref{61/1})
\begin{equation}\label{64/1}
\begin{array}{c}
{\bar D}^{\dot\alpha}F(x,\theta,\bar\theta)=0 \longrightarrow \, F=F(y,\theta),
\\[0.2cm]
\bar\nu_{\dot\alpha}\frac{\partial}{\partial x_{\alpha\dot\alpha}}F(y,\theta, \bar\nu)=0  
\longrightarrow \, F=F(\Xi_{\cal A})
\end{array}
\end{equation}
we get a chiral superfields $F(\Xi_{\cal A})$, depending on the $\theta$-twistors, to be their general solution 
because $\Xi_{\cal A}\equiv(-il_{\alpha},\bar\nu^{\dot\alpha}, \theta^{\alpha})$ is the general solution of the second constraint in Eqs. (\ref{64/1})
\begin{equation}\label{65/1}
\begin{array}{c}
\bar\nu_{\dot\alpha}\frac{\partial}{\partial x_{\alpha\dot\alpha}}\Xi_{\cal A}=0, 
\end{array}
\end{equation}
as it follows from Eqs. (\ref{7/1}) and (\ref{39/1}),  and  the relations  
\begin{equation}\label{66/1}
{\bar D}^{\dot\alpha}l_{\beta}=0, \quad 
\frac{\partial}{\partial y_{\alpha\dot\alpha}}l_{\beta}=\delta^{\alpha}_{\beta}\bar\nu_{\dot\alpha}.
\end{equation}
The both superfields $F(\bar Z^{\cal A})$ and $F(\Xi_{\cal A})$ describe  massless supermultiplets because they  obey the Klein-Gordon equations 
\begin{equation}\label{67/1}
\partial_{m}\partial^{m}  F(\bar Z)=0, \quad \partial_{m}\partial^{m} F(\Xi)=0, 
 \end{equation}
where $\partial_{m}\equiv(\sigma_{m})_{\dot\alpha\alpha}\partial^{\dot\alpha\alpha}
\equiv(\sigma_{m})_{\alpha\dot\alpha}\frac{\partial}{\partial x_{\alpha\dot\alpha}},\quad 
\partial^{\dot\alpha\alpha}=-\frac{1}{2}\tilde\sigma_{m}^{\dot\alpha\alpha}\partial^m$. 
Eqs. (\ref{67/1}) are satisfied due to the latter relations in (\ref{63/1}) and (\ref{66/1}) that 
provide  the vanishing multipliers $\nu^{\alpha}\nu_{\alpha}=\bar\nu^{\dot\alpha}\bar\nu_{\dot\alpha}=0$.

For the description of massless higher spin fields in supertwistor space the  contour integral
\begin{equation}\label{68/1}
G^{\alpha_{1}...\alpha_{2S}}(y,\theta)=\oint(d\nu^{\gamma}\nu_{\gamma})\nu^{\alpha_{1}}...\nu^{\alpha_{2S}}
 F(\nu^{\beta},i\nu^{\beta}y_{\beta\dot\beta}, 2\nu^{\beta}\theta_{\beta})
\end{equation}
generalizing the Penrose integral  \cite{PR} was used in \cite{Fbr}.
The superfunction $F(\bar Z)$ was supposed to be a generalized complex analytic function in the supertwistor space associated with the integral or half-integral spin $S$. To give meaning for (\ref{68/1}) as a  projective space integral  $F(\bar Z)$ was proposed to be homogenious of degree $-2(S-1)$ with a  $\nu$-contour enclosing singularities of $F$ for each fixed value of  $(y,\theta)$. 
 Next  one can expand $F(\bar Z)$ in a power series in $\eta$
\begin{equation}\label{69/1}
 F(\nu^{\beta},i\nu^{\beta}y_{\beta\dot\beta}, 2\nu^{\beta}\theta_{\beta})=
g_{0}(\nu^{\beta},i\nu^{\beta}y_{\beta\dot\beta})+ 2\nu^{\lambda}\theta_{\lambda}g_{1}(\nu^{\beta},i\nu^{\beta}y_{\beta\dot\beta})
\end{equation}
and then inserting Eq.  (\ref{68/1}) into  (\ref{68/1}) we get the component expansion for $G$  
\begin{equation}\label{70/1}
G^{\alpha_{1}...\alpha_{2S}}(y,\theta)=g_{0}^{\alpha_{1}...\alpha_{2S}}(y)+ 2\theta_{\lambda}g_{1}^{\alpha_{1}...\alpha_{2S}\lambda}(y), 
\end{equation}
where
\begin{equation}\label{71/1}
\begin{array}{c}
g_{0}^{\alpha_{1}...\alpha_{2S}}(y)=\oint(d\nu^{\gamma}\nu_{\gamma})\nu^{\alpha_{1}}...\nu^{\alpha_{2S}}g_{0}(\nu^{\beta},i\nu^{\beta}y_{\beta\dot\beta}),
\\[0.2cm]
g_{1}^{\alpha_{1}...\alpha_{2S}\lambda}(y)=\oint(d\nu^{\gamma}\nu_{\gamma})\nu^{\alpha_{1}}...\nu^{\alpha_{2S}}\nu^{\lambda}g_{1}(\nu^{\beta},i\nu^{\beta}y_{\beta\dot\beta}).
\end{array}
\end{equation}
We see that the component fields (\ref{71/1}) are totally symmetric in their spinor indices having the same chiralities. So, they  describe massless fields  with spin $S$  and  $S+\frac{1}{2}$,
because of  the zero mass  Klein-Gordon equations similar to  (\ref{67/1}). Moreover,  $g_{1}^{\alpha_{1}...\alpha_{2S}\beta}(y)$ satisfies to   the chiral Dirac equation 
\begin{equation}\label{72/1}
\varepsilon_{\alpha\alpha_{1}}\partial^{\dot\alpha\alpha}
g_{1}^{\alpha_{1}...\alpha_{2S}\lambda}(x)=0.
\end{equation}
It results in the absence of  the auxiliary field  together  with the $\theta^2\bar\theta^2\Box g_{0}^{\alpha_{1}...\alpha_{2S}}(x)$ and  
$\theta^{2}(\partial_{m}g_{1}^{\alpha_{1}...\alpha_{2S}}(x)\sigma^{m}\bar\theta)$ terms in the further expansion of  (\ref{70/1}) at the  point  $x_m$
\begin{equation}\label{72/2}
G^{\alpha_{1}...\alpha_{2S}}(y,\theta)=g_{0}^{\alpha_{1}...\alpha_{2S}}(x)+ 2\theta_{\lambda}g_{1}^{\alpha_{1}...\alpha_{2S}\lambda}(x) - i(\theta\sigma^{m}\bar\theta)g_{0}^{\alpha_{1}...\alpha_{2S}}(x). 
\end{equation}
So, the supersymmetry transformations of the component fields have the reduced form \cite{Fbr}
\begin{equation}\label{72/3}
\delta g_{0}^{\alpha_{1}...\alpha_{2S}}(x)=2\varepsilon_{\lambda}g_{1}^{\alpha_{1}...\alpha_{2S}\lambda}(x),\quad \quad
\delta g_{1}^{\alpha_{1}...\alpha_{2S}\lambda}(x)=2i\bar\varepsilon_{\dot\lambda}\partial^{\dot\lambda\lambda} g_{0}^{\alpha_{1}...\alpha_{2S}}(x)
\end{equation}
and close on the mass shell of  $g_{1}^{\alpha_{1}...\alpha_{2S}\lambda}$. 
For the case $S=0$  Eqs. (\ref{72/3}) coincide  with the transformation rules for the scalar or chiral supermultiplet \cite{WB} on the mass shell of the spinor field $g_{1}^{\lambda}$ and the vanishing auxiliary field. Thus, the  using  supertwistors imposes  rather severe constraints on the supermultiplet fields and gives rise a question whether it is possible to weaken these constraints. 
Below we shall give  a positive  answer for  this question based on using the above introduced  $\theta$-twistors  instead of the supertwistors. 

To  this  end let us  firstly note that the discussed  contour integral method may be also applied for complex analytic functions  in the $\Xi$-triple space
\begin{equation}\label{73/1}
 F(\Xi)\equiv F(-il_{\alpha},\bar\nu^{\dot\alpha}, \theta^{\alpha})=
f_{0}(-iy_{\beta\dot\beta}\bar\nu^{\dot\beta},  \bar\nu^{\dot\beta})-
 2\theta_{\lambda}f^{\lambda}(-iy_{\beta\dot\beta}\bar\nu^{\dot\beta}, \bar\nu^{\dot\beta}) +\theta^{2}f_{2}(-iy_{\beta\dot\beta}\bar\nu^{\dot\beta}, \bar\nu^{\dot\beta}),
\end{equation}
 where $ \theta^{2}\equiv\theta^{\gamma}\theta_{\gamma},\, \theta_{\alpha}\theta_{\beta}=\frac{1}{2}\varepsilon_{\alpha\beta} $.  Next we observe  that the supersymmetry generators (\ref{48/1}) in the  $\Xi$-space have the zero degree of gomogeneity. 
It means that the component fields $f_{i}$ in (\ref{73/1}) must have  the same degree of gomogeneity if they are accepted to be gomogenious functions. 
Then one can consider a contour integral
\begin{equation}\label{74/1}
\Phi^{\dot\alpha_{1}...\dot\alpha_{2S}}(y,\theta)=\oint(d\bar\nu^{\dot\gamma}\bar\nu_{\dot\gamma})\bar\nu^{\dot\alpha_{1}}...\bar\nu^{\dot\alpha_{2S}}
 F(\bar\nu^{\dot\beta},-i\bar\nu^{\dot\beta}y_{\beta\dot\beta}, \theta_{\beta})
\end{equation}
 similar  to the  supertwistor integral (\ref{68/1}), where $F(\Xi)$ is supposed to be a complex analytic function in the $\theta$-twistor space with the fixed degree of homogeneity $-2(S-1)$ and  a  $\bar\nu$-contour enclosing singularities of $F$ for each fixed point $(y,\theta)$. Inserting  (\ref{73/1}) into (\ref{74/1}) we get
\begin{equation}\label{75/1}
\begin{array}{c}
\Phi^{\dot\alpha_{1}...\dot\alpha_{2S}}(y,\theta)=f_{0}^{\dot\alpha_{1}...\dot\alpha_{2S}}(y)-2\theta_{\lambda}f^{\lambda\dot\alpha_{1}...\dot\alpha_{2S}}(y) +
\theta^{2}f_{2}^{\dot\alpha_{1}...\dot\alpha_{2S}}(y)
\end{array}
\end{equation}
where
\begin{equation}\label{76/1}
\begin{array}{c}
f_{0}^{\dot\alpha_{1}...\dot\alpha_{2S}}(y)=\oint(d\bar\nu^{\dot\gamma}\bar\nu_{\dot\gamma})\bar\nu^{\dot\alpha_{1}}...\bar\nu^{\dot\alpha_{2S}}f_{0}(-iy_{\beta\dot\beta}\bar\nu^{\dot\beta},  \bar\nu^{\dot\beta}),
\\[0.2cm]
f^{\lambda\dot\alpha_{1}...\dot\alpha_{2S}}(y)= \oint(d\bar\nu^{\dot\gamma}\bar\nu_{\dot\gamma})\bar\nu^{\dot\alpha_{1}}...\bar\nu^{\dot\alpha_{2S}}f^{\lambda}(-iy_{\beta\dot\beta}\bar\nu^{\dot\beta}, \bar\nu^{\dot\beta}),
\\[0.2cm]
f_{2}^{\dot\alpha_{1}...\dot\alpha_{2S}}(y)=\oint(d\bar\nu^{\dot\gamma}\bar\nu_{\dot\gamma})\bar\nu^{\dot\alpha_{1}}...\bar\nu^{\dot\alpha_{2S}}f_{2}(-iy_{\beta\dot\beta}\bar\nu^{\dot\beta},  \bar\nu^{\dot\beta}).
\end{array}
\end{equation}
Comparing the expansion  (\ref{76/1}) with (\ref{70/1}) we observe two new elements. The  first  of  them is  the  survival of the auxiliary field and the second is the  appearance of  the  spinor index
  $\lambda$ carrying chirality opposite to the  chiralities carried by the indices $\dot\alpha_{1}...\dot\alpha_{2S}$ in the spin $S+\frac{1}{2}$ field  $f^{\lambda\dot\alpha_{1}...\dot\alpha_{2S}}$. The chiral index  $\lambda$ in $f^{\lambda\dot\alpha_{1}...\dot\alpha_{2S}}$ has as its  origin the spinor $\theta_{\lambda}$ contrarily to the antichiral indices $\dot\alpha_{1},...,\dot\alpha_{2S}$ originating from the twistor component $\bar\nu_{\dot\alpha}$. 
 In view of this difference the Dirac equation similar to (\ref{72/1}) is satisfied only for the antichiral indices $\dot\alpha_{1},...,\dot\alpha_{2S}$
\begin{equation}\label{77/1}
\varepsilon_{\dot\alpha\dot\alpha_{1}}\partial^{\dot\alpha\alpha}
f^{\lambda\dot\alpha_{1}...\dot\alpha_{2S}}(x)=0,
\end{equation}
and doesn't affect on the supersymmetry transformation rules for the auxiliary fields. It is clear, because the component expansion of the chiral superfield $ \Phi^{\dot\alpha_{1}...\dot\alpha_{2S}}$ (\ref{75/1}) is going in $\theta_{\lambda}$ whose chirality is opposite to the $\bar\nu_{\dot\alpha}$ chirality.
To find the supersymmetry transformation law for the components of the superfield $\Phi^{\dot\alpha_{1}...\dot\alpha_{2S}}(y,\theta)$  (\ref{75/1})
\begin{equation}\label{78/1}
\begin{array}{c}
\delta\Phi^{\dot\alpha_{1}...\dot\alpha_{2S}}(y,\theta)\equiv
(\varepsilon_{\gamma}Q^{\gamma}+ 
\bar\varepsilon_{\dot\gamma}{\bar Q}^{\dot\gamma})\Phi^{\dot\alpha_{1}...\dot\alpha_{2S}}(y,\theta)=
\\[0.2cm]
\delta f_{0}^{\dot\alpha_{1}...\dot\alpha_{2S}}(y)-
2\theta_{\lambda}\delta f^{\lambda\dot\alpha_{1}...\dot\alpha_{2S}}(y) +
\theta^{2}\delta f_{2}^{\dot\alpha_{1}...\dot\alpha_{2S}}(y)
\end{array}
\end{equation}
we use the chiral representation (\ref{8/1}) for $Q^{\gamma}$ and ${\bar Q}^{\dot\gamma}$ and finally find 
\begin{equation}\label{79/1}
\begin{array}{c}
\delta f_{0}^{\dot\alpha_{1}...\dot\alpha_{2S}}(x)=-2\varepsilon_{\lambda}f^{\lambda\dot\alpha_{1}...\dot\alpha_{2S}}(x),
\\[0.2cm]
\delta f^{\lambda\dot\alpha_{1}...\dot\alpha_{2S}}(x)=2i\bar\varepsilon_{\dot\gamma}\partial^{\dot\gamma\lambda} f_{0}^{\dot\alpha_{1}...\dot\alpha_{2S}}(x)+ \varepsilon^{\lambda}f_{2}^{\dot\alpha_{1}...\dot\alpha_{2S}}(x),
\\[0.2cm]
\delta f_{2}^{\dot\alpha_{1}...\dot\alpha_{2S}}(x)=-4i\bar\varepsilon_{\dot\gamma}
\partial^{\dot\gamma\lambda} f_{\lambda}^{\dot\alpha_{1}...\dot\alpha_{2S}}(x).
\end{array}
\end{equation}
The further  expansion of $\Phi^{\dot\alpha_{1}...\dot\alpha_{2S}}(y,\theta)$ 
(\ref{75/1})  at the point $x_{m}$ is  given by 
\begin{equation}\label{80/1}
\begin{array}{c}
\Phi^{\dot\alpha_{1}...\dot\alpha_{2S}}(y,\theta)=f_{0}^{\dot\alpha_{1}...\dot\alpha_{2S}}(x)
-2\theta_{\lambda}f^{\lambda\dot\alpha_{1}...\dot\alpha_{2S}}(x) -
2i\theta_{\gamma}\bar\theta_{\dot\gamma}\partial^{\dot\gamma\gamma} f_{0}^{\dot\alpha_{1}...\dot\alpha_{2S}}(x)-
\\[0.2cm]
2i\theta^{2}\bar\theta_{\dot\gamma}\partial^{\dot\gamma\lambda}f_{\lambda}^{\dot\alpha_{1}...\dot\alpha_{2S}}(x)+\theta^{2}f_{2}^{\dot\alpha_{1}...\dot\alpha_{2S}}(x),
\end{array}
\end{equation}
where the term $\frac{1}{2}\theta^{2}{\bar\theta}^{2}\partial^{\dot\gamma\gamma}\partial_{\gamma\dot\gamma}f_{0}^{\dot\alpha_{1}...\dot\alpha_{2S}}(x)$ was droped because of the zero mass constraint 
(\ref{67/1})
\begin{equation}\label{81/1}
\begin{array}{c}
\Box\Phi^{\dot\alpha_{1}...\dot\alpha_{2S}}(y,\theta)=0 \longrightarrow \Box f_{0}^{\dot\alpha_{1}...\dot\alpha_{2S}}(x)=0.
\end{array}
\end{equation}

For sewing together of our results with the  known case of $S=0$ chiral supermultiplet it is suitable to use  notations  for  the  component  $f$-fields  similar to  used in \cite{WB} 
\begin{equation}\label{82/1}
\begin{array}{c}
f_{0}^{\dot\alpha_{1}...\dot\alpha_{2S}}={\sqrt{2}}A^{\dot\alpha_{1}...\dot\alpha_{2S}}
\equiv{\sqrt{2}}A^{...},
\quad
f_{2}^{\dot\alpha_{1}...\dot\alpha_{2S}}=\sqrt{2}F^{\dot\alpha_{1}...\dot\alpha_{2S}}
\equiv\sqrt{2}F^{...},
\\[0.2cm]
f_{\lambda}^{\dot\alpha_{1}...\dot\alpha_{2S}}=\psi_{\lambda}^{\dot\alpha_{1}...\dot\alpha_{2S}} 
\equiv\psi_{\lambda}^{...},
\end{array}
\end{equation}
where $(...)\equiv (\dot\alpha_{1}...\dot\alpha_{2S})$. 
Using these notations we present Eq. (\ref{80/1}) in a compact form 
\begin{equation}\label{83/1}
\begin{array}{c}
\frac{1}{\sqrt{2}}\Phi^{...}(y,\theta)= A^{...}(x) + i(\theta\sigma^{m}\bar\theta)\partial_{m} A^{...}(x) + \sqrt{2}\theta^{\lambda}\psi_{\lambda}^{...}(x) \,\, -
\\[0.2cm]
\frac{i}{\sqrt{2}}\theta^{2}(\partial_{m}\psi^{...}\sigma^{m}\bar\theta) + \theta^{2}F(x)
\end{array}
\end{equation}
which coincides with the massless chiral multiplet \cite{WB} if $S=0$. 
For  $S\neq0$  Eq. (\ref{83/1}) presents chiral supermultiplets of  massless higher spin fields with the particle content $$(\frac{1}{2},1),\, (1, \frac{3}{2}),\,(\frac{3}{2},2),\, ......., (S, S+\frac{1}{2})$$  accompanied by the  corresponding  auxiliary fields for any of chosen spin $ S=\frac{1}{2},1, \frac{3}{2},2,\frac{3}{2}, ....$ from the infinite integral/half spin chain.   
The transformation rules (\ref{79/1}) for the higher spin multiplets  (\ref{83/1}) rewritten 
 in the notations (\ref{82/1}) are 
\begin{equation}\label{84/1}
\begin{array}{c}
\delta A^{...}=\sqrt{2}\varepsilon^{\lambda}\psi_{\lambda}^{...},
\\[0.2cm]
\delta\psi_{\lambda}^{...}=i\sqrt{2}(\sigma_{m}\bar\varepsilon)_{\lambda}\partial^{m}A^{...}+ 
\sqrt{2}\varepsilon_{\lambda}F^{...},
\\[0.2cm]
\delta F^{...}=i\sqrt{2}(\bar\varepsilon{\tilde\sigma}_{m}
\partial^{m}\psi^{...})
\end{array}
\end{equation}
 and coincide  with the transformation rules for the $S=0$ chiral multiplet of the weight 
$n=\frac{1}{2}$  \cite{WB}  if we assume  $A^{...}=A,\, F^{...}=F$ and $\psi_{\lambda}^{...}= \psi_{\lambda}$  in the  relations (\ref{84/1}). 

 As it was  above shown the $\theta$-twistor superspace is invariant under the axial rotations (\ref{55/1}) and  thus one can consider these phase transformations as inducing the  $R$-symmetry transformations for the superfield $F(\Xi)$ (\ref{73/1})
\begin{equation}\label{85/1}
\begin{array}{c}
 F'(-il_{\alpha}, \bar\nu^{\dot\alpha},  e^{i\varphi}\theta^{\alpha})=e^{2in\varphi}F(-il_{\alpha}, \bar\nu^{\dot\alpha}, \theta^{\alpha}), 
\\[0.2cm]
\delta F(\Xi)=2in\delta\varphi F(\Xi)-\delta\varphi U_{5} F(\Xi),
\end{array}
\end{equation}
where $n$ is the correspondent $R$ number. Then  taking into account (\ref{74/1}) we get 
  the $R$-symmetry transformation of $\Phi^{\dot\alpha_{1}...\dot\alpha_{2S}}(y,\theta)$
\begin{equation}\label{86/1}
\Phi'^{\dot\alpha_{1}...\dot\alpha_{2S}}(y,\theta)=
e^{2in\varphi}\Phi^{\dot\alpha_{1}...\dot\alpha_{2S}}(y,e^{-i\varphi}\theta),
\end{equation}
 rezulting in the $R$-symmetry transformation  for the component fields similar  to \cite{WB}
\begin{equation}\label{87/1}
\begin{array}{c}
{f_{0}}'^{\dot\alpha_{1}...\dot\alpha_{2S}}(x)=
e^{2in\varphi}f_{0}^{\dot\alpha_{1}...\dot\alpha_{2S}}(x),
\quad
{f_{2}}'^{\dot\alpha_{1}...\dot\alpha_{2S}}(x)=
e^{2i(n-1)\varphi}f_{2}^{\dot\alpha_{1}...\dot\alpha_{2S}}(x),
\\[0.2cm]
{f'}_{\lambda}^{\dot\alpha_{1}...\dot\alpha_{2S}}(x)=
e^{2i(n-\frac{1}{2})\varphi}f_{\lambda}^{\dot\alpha_{1}...\dot\alpha_{2S}}(x).
\end{array}
\end{equation}
Having defined the $R$-symmetry transformations for the chiral superfields (\ref{83/1}) one can construct supersymmetric Langrangians by analogy with  the case of scalar multiplet \cite{WB} and to expect 
 these Lagrangians to be renormalizable in view of the $R$-symmetry.

\section{Conclusion}

The supertwistor conception based on using the supersymmetric chiral Cartan form was disscussed 
resulting in the construction of a new type of supersymmetric twistor called $\theta$-twistor versus the supertwistor. The $\theta$-twistor was defined as a triple including two commuting spinors supplemented by the anticommuting spinor $\theta$ to form a non-linear supersymmetry representation. 
We revealed that the $\theta$-twistors and supertwistors appear as the general solutions of two different supersymmetric and Lorentz covariant constraints admissible in the chiral superspace extended by the Penrose spinors $\nu$ and $\bar\nu$. 
 The symmetry properties of the $\theta$-twistor superspace were studied and its closure under the superconformal group, except the (super)conformal boosts, was established. Using the Penrose contour integral the known chiral superfields depending only on $x$ and $\theta $ were restored and their generalization to  higher spins was found. 
The new  superfields describe an  infinite chain of massless higher spin chiral supermultiplets $(\frac{1}{2},1),\, (1, \frac{3}{2}),\,(\frac{3}{2},2),\, ......., (S, S+\frac{1}{2})$ generalizing the scalar supermultiplet $(0,\frac{1}{2})$. 
These supermultiplets contain auxiliary fields which are absent in the supertwistor description and their supersymmetry transformations generalize  the known transformations of the scalar supermultiplet $(0,\frac{1}{2})$. Unlike the supertwistor description, where all superfields carry only chiral (or antichiral) spinor indexes, the introduced here superfields carry both the chiral and antichiral indexes.      
For simplicity the $D=4 \, N=1$ supersymmetry case was studied here. However, the construction has a straightforward generalization to more general  case of the $SU(N)$ internal symmetry and higher 
$D=2,3,4(mod8)$, allowing the  Majorana spinors, by using the change: 
$(\theta_{\alpha},\,{\bar\theta}_{\dot\alpha})\rightarrow \theta^{i}_{a}$, where $a$ is the Majorana spinor index  and $i \in SU(N)$. 
Moreover, the $\theta$-twistor conception  is naturally generalized for the tensorial superspaces used for the on-shell description of higher spin field (see e.g. \cite{VBLS} and refs. there).
The generalizations as well as studying Langrangians built from the considered  superfields will be considered in other place.

\section{Acknowledgements}

The author thanks Fysikum at the Stockholm University for kind hospitality and I. Bengtsson, F. Hassan,   and U. Lindstrom who pointed out the refs.  \cite{Sieg},  for useful discussions. The work was partially supported by the grant of the Royal Swedish Academy of Sciences.

\end{document}